\documentclass[aps,prx,twocolumn,superscriptaddress,citeautoscript]{revtex4-1}

\usepackage{amsmath,amssymb}
\usepackage{bm}
\usepackage{graphicx}
\usepackage{epstopdf}
\usepackage{latexsym}
\usepackage{subfigure}
\usepackage{color}
\usepackage{dsfont} 
\usepackage{wasysym} 
\usepackage{hyperref} 

\definecolor{airforceblue}{rgb}{0.36, 0.54, 0.66}

\newcommand{\be}{\begin{equation}}
\newcommand{\ee}{\end{equation}}
\newcommand{\bea}{\begin{eqnarray}}
\newcommand{\eea}{\end{eqnarray}}

\begin{document}

\title{Disorder-induced entanglement in spin ice pyrochlores}

\author{Lucile Savary}
\email{author to whom correspondence should be addressed:    \\savary@mit.edu}
\affiliation{Department of Physics, 
  Massachusetts Institute of Technology, 77 Massachusetts Ave.,
  Cambridge, MA 02139}
\author{Leon Balents}
\affiliation{Kavli Institute for Theoretical Physics, University of
  California, Santa Barbara, CA 93106-4030}

\date{\today}
\begin{abstract}
 We propose that in a certain class of magnetic materials, known as non-Kramers `spin ice,' disorder induces quantum entanglement. Instead of driving glassy behavior, disorder provokes quantum superpositions of spins throughout the system, and engenders an associated emergent gauge structure and set of fractional excitations. More precisely, disorder transforms a classical phase governed by a large entropy, classical spin ice, into a quantum spin liquid governed by entanglement. As the degree of disorder is increased, the system transitions between (i) a ``regular'' Coulombic spin liquid, (ii) a phase known as ``Mott glass,'' which contains rare gapless regions in real space, but whose behavior on long length scales is only modified quantitatively, and (iii) a true glassy phase for random distributions with large width or large mean amplitude. 

\end{abstract}

\maketitle

Entanglement, the extent to which measurement of one subsystem affects
the state of another, is an essential non-classical feature of quantum
mechanics.  While entanglement has been achieved and controlled for
small numbers of quantum bits (``qubits''), {\em many-body}
entanglement of a thermodynamically large system is an exciting
frontier \cite{amico2008entanglement,chen2010local}.  Long range
entanglement engenders exotic phenomena such as fractional quantum
numbers and emergent topological excitations, and is important not
only in the realm of materials but even in the theory of fundamental
forces: light and gravity themselves may emerge from underlying
quantum entanglement \cite{swingle2014universality}.  Theoretically,
the exemplars of such massive ``long range'' entanglement are Quantum
Spin Liquids (QSLs), states of quantum magnets in which electronic
spins reside in macroscopic superpositions of infinitely many
microstates \cite{balents2010spin}.  QSLs are actively sought in quantum magnets with strongly frustrated interactions that discourage the freezing of electronic moments into an ordered pattern, which is the enemy of entanglement.  The strategy has been to seek materials which in their ideal, perfect form, accidentally host particular spin interactions that give way to a QSL ground state.  However, these QSLs are typically fragile, and the inevitable and uncontrollable deviations of a real material from the ideal, consisting of additional interactions and/or random disorder, can remove or modify the QSL essentially, and at the very least muddle the interpretation of experiments. In many frustrated magnets disorder in particular leads to glassy freezing which overwhelms entanglement.

\begin{figure}[htbp]
\label{fig:phasediagram}
\includegraphics[width=\linewidth]{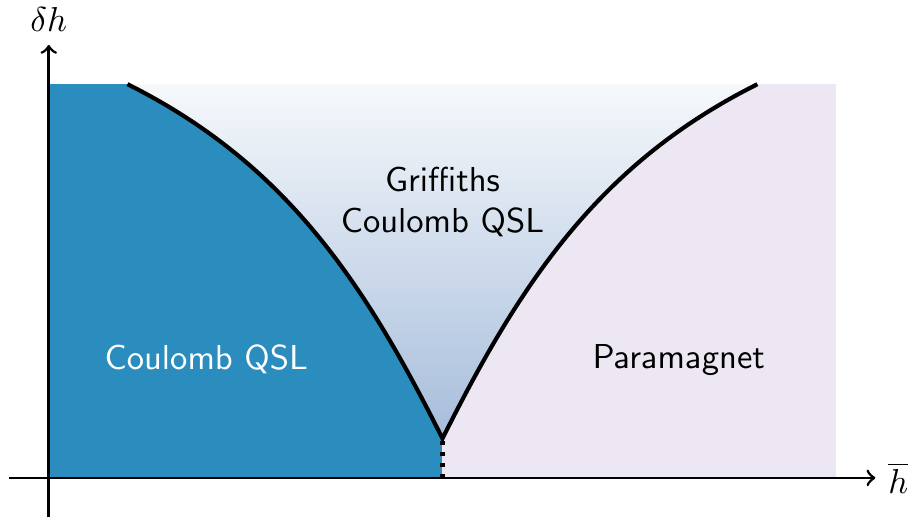}
  \caption{Phase diagram in the mean strength of disorder $\overline{h}$ --- disorder $\delta h$ plane. The dotted line indicates a first order transition, while the solid lines represent second order transitions or crossover (between the Coulomb QSL and Griffiths Coulomb QSL). In the disordered boson language, the paramagnet is a ``superfluid'' (Higgs) phase, the Griffiths phase is a Mott glass, and the Coulomb QSL is a ``Mott insulator''.}
\end{figure}

Here we take a different strategy and propose to {\em use} the
disorder itself to generate long-range entanglement.  Because disorder
is not intrinsic, it can be readily tuned so that serendipity is no
longer required to find the QSL state.  The key theoretical
observation is that, from the most general point of view, disorder is
simply the breaking of translational symmetry, and one of the
essential features of long range entanglement is that it is completely
independent of any symmetry.  The well-known necessity of disorder to
describe the Quantum Hall Effect illustrates this fact. Hence there is
no fundamental obstacle to a QSL state in a strongly disordered
system.  Yet it is far from obvious how to actually realize such a
``random QSL''.  Here we show that the essential ingredients are
present in spin ice materials \cite{gingras11:_spin_ice} such as
Ho$_2$Ti$_2$O$_7$ and Pr$_2$Zr$_2$O$_7$ with {\em non-Kramers}
magnetic ions.  We construct a model for disorder in these materials,
and show that it indeed supports not one but two QSL phases, one of
which is a long range entangled analog of the ``Mott glass'' phase of
disordered bosons \cite{altman2004phase,weichman2008dirty}.   On
application of a physical magnetic field we obtain an even more glassy
``Bose glass'' QSL phase \cite{fisher1989boson}.  We emphasize these
are true QSLs with long range entanglement, emergent gauge structure,
and exotic non-local excitations.   The glassy QSLs differ from the
pure QSL by having additional gapless but localized excitations at low
energy.    To our knowledge this is the first proven example where
true QSL states are engendered by disorder.  We emphasize that our
model applies to the archetypal {\em classical} spin ice material
Ho$_2$Ti$_2$O$_7$, and predicts that it can be tuned to a {\em
  quantum} spin liquid by controlled introduction of disorder.  The
full phase diagram is shown in Fig.~1.

Our analysis begins with the atomic physics of trivalent rare earth
ions in the spin ice pyrochlores \cite{gardner2010magnetic}.  Many of these -- e.g. Ho$^{3+}$, Tb$^{3+}$, Pr$^{3+}$ -- are non-Kramers ions, with an even number of electrons, and thus are not guaranteed to have a degeneracy by time-reversal symmetry.  Instead, the low energy levels of these ions comprise an isolated doublet whose degeneracy is protected by the local $D_{3d}$ point group symmetry.  If the doublet is well-separated from higher states, as it is in Ho$^{3+}$ and Pr$^{3+}$, the entire description of the magnetism of these materials can be represented by a pseudo-spin 1/2 operator, $\vec{{\sf S}}_i$ for each rare earth site.  The non-Kramers nature of the ion implies that, under time-reversal symmetry, in the local basis aligned with the $\langle 111\rangle$ axis of the site, the ``up'' and ``down'' spin levels interchange: i.e.\  $\hat{\Theta}|\pm\frac{1}{2}\rangle =  |\mp \frac{1}{2}\rangle$, where $\hat{\Theta}$ is the anti-unitary time-reversal operator.  Note the absence of a minus sign in this relation.  It implies that  $\hat{\Theta}^2 = +1$, which defines the non-Kramers case.  Following from this, one may see that the spin operator transforms according to ${\sf S}_i^z \rightarrow - {\sf S}^z_i$ under time-reversal, while ${\sf S}_i^x$ and ${\sf S}_i^y$ are time-reversal invariant.  

The Hamiltonian is a sum of zero field terms and the Zeeman coupling of the magnetic moment operator to an applied magnetic field.  In clean spin ice systems, an excellent first approximation is given by the nearest-neighbor spin ice Hamiltonian,
\begin{equation}
  \label{eq:1}
  H_0 = J \sum_{\langle ij\rangle} {\sf S}_i^z {\sf S}_j^z - {\bf B} \cdot \sum_i g {\sf S}_i^z \hat{\bf e}_i .
\end{equation}
The first term, with $J>0$, is a frustrated Ising interaction between
spins.  It appears antiferromagnetic in the local basis but represents
ferromagnetic coupling of the magnetic moments in a global frame.  The
second term is the only symmetry-allowed interaction of the magnetic
field with the spins in the non-Kramers case: the magnetic moment
operator is, by symmetry, ${\bf m}_i = g {\sf S}_i^z \hat{\bf e}_i $.
In principle there is, in addition, a long-range dipole interaction
between moments.  It has been shown that this can be largely subsumed
into the ``pseudo-dipolar'' $J$ term above \cite{isakov2005spin}, so
we neglect it in the following.   Quantum exchanges coupling in-plane
components ${\sf S}_i^x, {\sf S}_i^y$ on nearest-neighbor sites can
also occur, but are small in Ising-like systems.  For example, in
Pr$^{3+}$, it is estimated that the probability to be in the maximal
$j_z=\pm 4$ states of the $j=4$ levels is 93\% \cite{kimura2013quantum}, while Ho$^{3+}$, which has $j=8$, is even more Ising-like.  

Now we examine the effect of disorder.  We consider {\em non-magnetic} disorder on the Ti or O sublattices, so that no spins are added or removed from the system, and assume there is no ordered Jahn-Teller distorsion as appears to be the case in experiments.  Rather, disorder generates (electrostatic) crystal fields which lower the symmetry of the rare earth site, and hence can split the non-Kramers doublet.  Due to time-reversal symmetry, these crystal fields couple directly to the in-plane components of $\vec{{\sf S}}_i$.  Hence disorder adds the term
\begin{equation}
  \label{eq:2}
  H'  = - \frac{1}{2}\sum_i \left( \eta^*_i {\sf S}_i^- + \eta_i {\sf S}_i^+\right),
\end{equation}
where $\eta_i$ is a random complex number, acting as an XY ``random field'' (though we caution there is no true field, and $H'$ is time-reversal invariant).  In general, the problem is specified by giving the full distribution of the random fields, $P[\{ \eta_i \}]$, and the statistical space group symmetry of the crystal should be respected by this distribution.  We will largely focus on the simplest limit of independent, identically distributed random variables, i.e.\ $P[\{ \eta_i \}] = \prod_i p(\eta_i)$ (but this is not essential).

The full Hamiltonian, $H=H_0+H'$, defines a quenched random transverse field Ising model.  It can be simplied by defining $\eta_i = h_i e^{i\alpha_i}$, where $h_i>0$ is real and $0\leq \alpha_i <2\pi$.  The phase $\alpha_i$ can be removed by a basis rotation around the local $z$ axis, generated by the unitary operator $U=\prod_i e^{i\alpha_i {\sf S}_i^z}$.  After the transformation, we have
  \begin{equation}
    \label{eq:3}
    H \rightarrow U^\dagger HU = J \sum_{\langle ij\rangle} {\sf S}_i^z {\sf S}_j^z - \sum_i h_i {\sf S}_i^x - {\bf B} \cdot \sum_i g {\sf S}_i^z \hat{\bf e}_i.
  \end{equation}
We see that in zero applied field, ${\bf B}=\mathbf{0}$, this is really the standard transverse field Ising antiferromagnetic model, with random magnitudes of the transverse field, drawn from some distribution $p(h)$.  We expect that a variety of distributions can be tuned experimentally.  

{\em \underline{\textbf{Perturbative regime: $h_i \ll J$.---}}} When all or nearly all the $h_i \ll J$, (i.e.\ the probability that $h> fJ$, with $f$ a small fraction of 1, is small: $\int_{fJ}^\infty p(h) dh \ll 1$) we may apply perturbation theory.  We obtain the effective Hamiltonian at {\em sixth} order in the transverse fields within the degenerate manifold of classical spin ice states (this is a non-trivial exercise which must be carried out for arbitrary site-dependent fields $h_i$ -- see Supp. Mat.):
\begin{equation}
  \label{eq:4}
  H_{\rm eff} = - \sum_{\hexagon} \left( K_{ijklmn} {\sf S}_{i}^+{\sf S}_{j}^- {\sf S}_{k}^+{\sf S}_{l}^- {\sf S}_{m}^+{\sf S}_{n}^- + {\rm h.c.} \right),
\end{equation}
where
\begin{equation}
  \label{eq:5}
  K_{ijklmn} = \frac{63h_ih_jh_kh_lh_mh_n}{16J^5}.
\end{equation}
Eqs.~(\ref{eq:4},\ref{eq:5}) define a {\em random ring exchange}
model. As shown first by Hermele {\em et al.}\
\cite{hermele2004pyrochlore}, when $K$ is constant, the ring exchange
model has the structure of a compact U(1) gauge theory, in which ${\sf
  S}_i^\pm$ plays the role of a U(1) gauge connection (exponential of
a gauge field) on the links of the dual diamond lattice formed from
the tetrahedron centers, and ${\sf S}_i^z$ acts as the conjugate
``electric'' field.   On general grounds, such a theory can support a
trivial ``confined'' phase which is short range entangled and a
deconfined Coulomb phase, which is long range entangled
\cite{banks1977phase}.  In the latter, the compactness is unimportant
and the low energy physics is an emergent quantum electrodynamics,
with a gapless photon and gapped electric and magnetic charged
quasi-particles.   This is a U(1) QSL phase.  Numerical studies have
shown that the ground state of this specific model for constant $K$ is
in the U(1) QSL phase \cite{banerjee2008unusual,shannon2012quantum,kato2015}.  

{\em \textbf{Weak randomness: $\delta h \ll \overline{h}$.---}} Let us now consider first {\em weak} randomness, i.e.\ a distribution $p(h)$ peaked around $\overline{h}$ with small width $\delta h \ll \overline{h}$.  The obvious potential instabilities of the $U(1)$ QSL phase are due to vanishing gaps for electric and magnetic charges.  The electric charges (in standard quantum conventions) correspond to tetrahedra violating the ice rules, and in the perturbative limit have a gap of order $J\gg K$, and hence remain gapped regardless of the distribution $p(h)$.  The magnetic charges have a gap of order $\overline{K} \sim \overline{h}^6/J^5$, which is still much larger than the random perturbation to $H_{\rm eff}$ which is of order $\delta K \sim (\overline{h}/J)^5 \delta h$.  Thus the gap to magnetic charges is also robust.  

What of the photon? Due to the absence of magnetic charges, the low energy effective theory is a non-compact U(1) gauge theory.  In the continuum limit, the most general allowed Hamiltonian including disorder takes the form, dictated by gauge invariance and time-reversal symmetry:
\begin{equation}
  \label{eq:6}
  H_{\rm photon} = \int\! d^3x \Big\{ \frac{\epsilon}{2}(1+v_E(x)) |E|^2 + \frac{1}{2\mu}(1+v_B(x))|B|^2 \Big\},
\end{equation}
where $v_E(x)$ and $v_B(x)$ are zero-mean random functions of space,
and $\epsilon,\mu$ are the effective dielectric constant and magnetic
permeability, respectively.  Simple power-counting shows that both
random terms are strongly {\em irrelevant} at low energy and long
distances (with short-range correlations, $[v] = L^{-3/2}$ in three
dimensions).  The key point is that gauge invariance forces disorder
only to couple to $E$ and gradients of the vector potential $A$, so
that, even if we relax the constraint of time-reversal symmetry in
Eq.~(\ref{eq:6}), the photon remains stable.  This is similar to the
suppression of scattering of acoustic phonons at low energy in a
disordered crystal \cite{PhysRevB.27.5592}, and the lack of
localization of light in a disordered photonic material at low
frequency \cite{john1991localization}.

{\em \textbf{Larger disorder: $\delta h\sim\overline{h}$.---}} We have
established the stability of the U(1) QSL with weak disorder.  Now let
us consider increasing the disorder, still within the perturbative
regime, i.e. the random ring model with $\delta h \sim \overline{h}$.
In general the ground state depends now on the full distribution
$p(h)$ (or the induced distribution $p(K)$).  The gap to electric
charges remains robust, but the magnetic gap may close, leading to
confinement.  The physical mechanism whereby this might occur is
order-by-disorder \cite{villain1980order,henley1989ordering}.  The ring Hamiltonian, Eq.~(\ref{eq:4}), is a kind of ``hopping'' in the high dimensional manifold of classical spin ice states.  One outcome, which is realized for the uniform case, is that the ground state is delocalized across an extensive subset of this manifold: this is the QSL state, which is a massive superposition of classical states.  Such a state obtains the same energy for each ring term.  We can also imagine a different state which gets a lower energy for some ``strongly resonating'' ring terms (better than the delocalized state) but sacrifices energy for other rings -- in the non-random case this necessarily breaks lattice symmetries.  The fact that the ground state is a uniform QSL for constant $K$ implies that the energy sacrificed by non-resonating hexagons outweighs the energy lowering of the resonating ones in the competing confined state in that case.   However, this energy balance is tilted as disorder is increased.  By choosing the hexagons with larger $K$ to resonate, the order-by-disorder state becomes more competitive.  It is quite non-trivial to construct such a state since the strong hexagons are random, and leave behind many spins that do not participate in such hexagons.  What these spins do is subtle and the frustration associated with their indecision likely tends to stabilize the more uniform QSL state.  Nevertheless, we expect that such a confined state may occur when $p(h)$ is sufficiently broad, though this depends in a non-universal way on the full distribution.  It is natural to think that the confinement transition to such a state has a dual interpretation as condensation of the magnetic charged excitations of the QSL phase.  

{\em \underline{\textbf{Non-perturbative case: $h_i \sim J$.---}}}  When the transverse fields are not small, the perturbative treatment no longer applies.  Instead, we adopt the slave rotor representation introduced for the uniform quantum spin ice problem in Ref.~\onlinecite{savary2012coulombic}, and discuss the full phase diagram in this framework.  This is an exact rewriting of the original spin system, by introducing explicit operators to track spinons (or electric charges) on the sites $a,b,\cdots$ of the diamond lattice.  The charge is $Q_a = \epsilon_a \sum_{i \in a} {\sf S}_i^z$, where $\epsilon_a = +1(-1)$ on the diamond A (B) sublattice.  A conjugate phase $\varphi_a$ is defined by $[\varphi_a,Q_b]=i\delta_{ab}$.  Then the spin operators are rewritten as ${\sf S}_i^z = {\sf s}_{ab}^z$ and ${\sf S}_i^+ =\Phi_a^\dagger {\sf s}_{ab}^+ \Phi_b^{\vphantom\dagger}$, and $h_i=h_{ab}$, where $a,b$ are the two tetrahedra sharing site $i$, on the A and B sublattices, respectively, and $\Phi_a = e^{-i\varphi_a}$.  The ${\sf s}_{ab}^\mu$ spins are canonical spin-1/2 degrees of freedom, and for convenience we define ${\sf s}_{ba}^z = - {\sf s}_{ab}^z$ and ${\sf s}_{ba}^\pm = {\sf s}_{ab}^\mp$.  Then the Hamiltonian, Eq.~(\ref{eq:3}) becomes
\begin{equation}
  \label{eq:7}
  H = \frac{J}{2} \sum_a Q_a^2 - \frac{1}{2}\sum_{\langle ab\rangle} h_{ab} \left[ \Phi_a^\dagger {\sf s}_{ab}^+ \Phi_b^{\vphantom\dagger}+{\rm h.c.}\right].
\end{equation}
Like in the uniform quantum spin ice problem, this Hamiltonian contains a potential term, and a kinetic term which represents electric charges (spinons) hopping on top of a fluctuating background gauge field. This kinetic term  appears only for non-zero ``fields,'' here disorder. The coupling of the spinons to the gauge field leads to a strongly-interacting problem.

{\em \textbf{Gauge Mean Field Theory: no gauge field fluctuations.---}} First, we discuss an approximate solution obtained by gauge Mean Field Theory (gMFT) \cite{savary2012coulombic}, which, in the present case essentially consists in suppressing the fluctuations of the gauge field. Namely, we perform the replacement $\Phi^\dagger\mathsf{s}\Phi\rightarrow\Phi^\dagger\Phi\langle\mathsf{s}\rangle+\langle\Phi^\dagger\Phi\rangle\mathsf{s}-\langle\Phi^\dagger\Phi\rangle\langle\mathsf{s}\rangle$. The resulting mean field Hamiltonian is composed of two decoupled parts, a ``spin'' $\vec{\mathsf{s}}$ in a random field, and a quadratic spinon hopping Hamiltonian:
\begin{eqnarray}
  \label{eq:8}
  H_\Phi&=&\frac{J}{2} \sum_a Q_a^2 -\frac{1}{2}\sum_{\langle
            ab\rangle}\left[t_{ab}\Phi^\dagger_a\Phi_b^{\vphantom{\dagger}}+{\rm
            h.c.}\right]\\
 &=& \frac{J}{2} \sum_a Q_a^2 - \sum_{\langle ab\rangle} t_{ab} \cos (\varphi_a-\varphi_b),
\end{eqnarray}
with $t_{ab}=h_{ab}\langle\mathsf{s}^+_{ab}\rangle$, which we assumed to be real in the right-hand side expression, as is indeed the case for the gMFT solution. We recognize this as the Hamiltonian of a (three-dimensional) array of Josephson junctions, i.e. a quantum XY/rotor model, coupling ``grains'' on the diamond lattice with random Josephson coupling $t_{ab}$.

{\em Uniform field  ---} While our primary interest is in disorder, we
first consider the case of a uniform $h$, for which Eq.~(\ref{eq:7})
is translationally invariant, and so are the mean field Hamiltonians,
and we make the Ansatz that $\langle\mathsf{s}\rangle$ (and hence
$t_{ab}$) be also uniform. Then the quantum XY model in
Eq.~(\ref{eq:8}) is expected to have two phases: a ``superfluid''
state with $\langle e^{i\varphi_a}\rangle \neq 0$ and a Mott insulator
phase with $\langle e^{i\varphi_a}\rangle=0$ and a gap to all
excitations.  The ``superfluid'' state corresponds to the Higgs phase
of the gauge theory -- the trivial transverse polarized state of the
original model.  In general the precise location of the transition
between these phases requires a quantum Monte Carlo calculation.
However, we can obtain an analytical approximation, widely used for
such rotor models, by making a ``spherical approximation'' which
replaces the constraint  $\Phi^\dagger_a\Phi^{\vphantom{\dagger}}_a=1$
by its average, implemented by a Lagrange multiplier $\lambda$ (see
Appendix~B). 
Within this approach, we immediately find that the Higgs transition, where the spinons become gapless, between the Mott and superfluid states takes place at $(h/J)_c\approx0.35$.  Below $(h/J)_c$ the system is in the Coulomb phase (``Mott''), and characterized within gMFT by $\langle\Phi\rangle=0$.  In this phase, fluctuations around the mean field solution reproduce the photon Hamiltonian, c.f.\ Eq.~(\ref{eq:6}).  We expect that the transition to the paramagnetic phase in the disordered case occurs at a similar magnitude of $\overline{h}/J$.

{\em Random field.---} Now we return to the full problem with random
$h_{ab}$, hence random $t_{ab}$.  The gMFT Hamiltonian in
Eq.~(\ref{eq:8}) then describes a well-studied ``dirty boson''
problem, notably with {\em particle-hole symmetry} $Q_a\rightarrow
-Q_a$, $\varphi_a \rightarrow - \varphi_a$.  We can trace this back to
the time-reversal symmetry of the original model.  Due to disorder, an
additional phase emerges between the Mott insulator and superfluid: a
gapless insulating state which has been called a ``Mott glass''
\cite{altman2004phase,weichman2008dirty}.  We recapitulate its
description here.  In most respects the Mott glass is similar to the
Mott insulator (the Coulomb phase in spin language), but differs by
the presence of rare regions which look locally superfluid (trivial,
paramagnetic, polarized), and consequently have very small gaps
controlled by their finite size.  In an infinite system, arbitrarily
large regions of this type can be found, leading to a vanishing gap in
the thermodynamic sense.  The situation, in which some non-local
quantities, e.g.\  gaps, are dominated by rare regions, is known as a
``Griffiths phase'' \cite{griffiths1969}.  Due to particle-hole
symmetry, the superfluid regions are exceedingly rare, and numerics
suggest \cite{iyer2012mott} that the superfluid clusters are exponentially distributed in their size, i.e.\ the density of superfluid regions of $s$ sites decays as $e^{-(s/s_0)^\gamma}$, with $\gamma \approx 1$ and $s_0$ a constant.  This in turn implies that the largest superfluid cluster in a system of size $L$ grows logarithmically, $s_{\rm max} \sim \ln L$.  Since a superfluid region of size $s$ has a gap of order $1/s$, the finite-size gap of the Mott glass is therefore order $\Delta_L \sim 1/\ln L$, which vanishes in the thermodynamic limit.  

\begin{figure}[htbp]
\includegraphics[width=\linewidth]{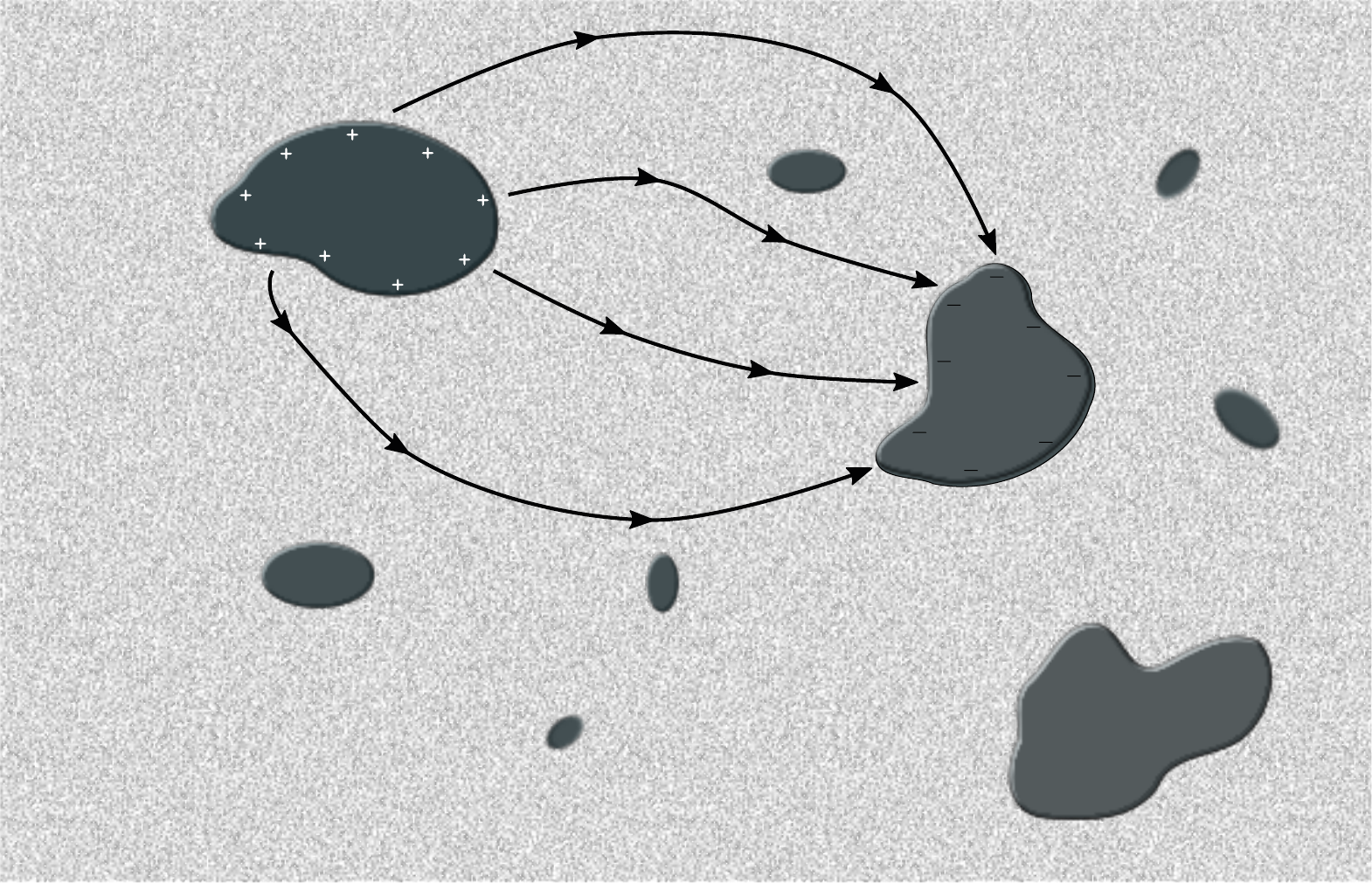}
  \caption{Pictorial representation of the Griffiths Coulomb QSL phase. The black blobs represent puddles (inclusions) of the paramagnetic (superfluid) phase where the electric excitations have small gaps controlled by the finite size of the puddles (they are gapless in the thermodynamic limit).   Rare large puddles can be ``charged'' with emergent electric excitations with very low energy, generating flux lines as shown on the figure.}
  \label{fig:griffiths}
\end{figure}

{\em \textbf{Beyond gMFT.---}} These properties need to be translated
into their {\em physical} consequences for the full problem, beyond
gMFT.  A picture (Fig.~2) is as follows: fluctuations convert the
Mott insulating bulk with superfluid inclusions to a Coulombic bulk
with Higgs inclusions -- a Coulombic Griffiths phase.  This is
analogous simply to a dielectric medium with embedded superconducting
grains \cite{deutscher2006}.  The latter exclude the gauge fields and act as low energy hosts for electric charges -- specifically, the ``charging energy'' for a grain of linear size $\ell$ is of order $1/\ell$ (note this is larger than in the Mott glass without gauge fluctuations, due to the Coulomb interaction mediated by the emergent gauge field).  Since the electric charges correspond to spinons, the spinon gap vanishes, with the gapless low energy spinon states localized on these grains.  Modifications of the photon are like those of an electromagnetic wave in a dielectric+superconducting ``metamaterial''.  Such waves are {\em insensitive} to rare regions, but are dominated by typical ones.  At low frequency the system behaves as an effective medium with an enhanced dielectric constant, but at frequency and wavevector scales comparable to the gap and inverse typical size/spacing of the Higgs regions, the photon will scatter and develop an intrinsic, disorder-dominated, linewidth.  Since the emergent photon continues to remain gapless and propagate, and electric and magnetic charges can still exist in isolation in the system, the Griffiths phase should still be considered a Coulombic spin liquid.  

In the full treatment, the mean-field superfluid phase, as for the pure system, becomes the confined paramagnetic phase, with no long-range entanglement.  For weak disorder, i.e. $\delta h \ll \overline{h}$, there can be a true gapped paramagnet, but for strong disorder we expect a zero gap state with localized low energy excitations -- a Griffiths paramagnet.  So the zero temperature phase diagram contains both the usual Coulombic liquid with gapped electric and magnetic charges, a Griffiths Coulomb liquid with gapless electric charges, and the thermally insulating unentangled paramagnetic state.  It is worth noting that the application of a physical magnetic field (which couples to ${\sf S}_i^z$ rather than ${\sf S}_i^\pm$) breaks time-reversal symmetry and hence the particle-hole symmetry of the emergent gauge theory.  Consequently, we would expect it to convert the Mott glass to a Bose glass state, which has much stronger Griffiths effects.  The experimental ramifications would be a an excellent subject for future research.

{\em \underline{\textbf{Phase transitions.---}}} Disorder has major
effects upon the transition from the QSL to the trivial state.  In
mean field theory and without disorder, the transition is described by
condensation of a complex field representing the spinon or Higgs
field.  This must be corrected by both disorder and coupling to the
$U(1)$ gauge field, effects which have been considered separately
before but not together in the literature.   The gauge coupling alone
renders this a U(1) abelian Higgs transition, governed by an effective
action which has the form of a Ginzburg-Landau theory.  The coupling
to the gauge field is marginal in the Renormalization Group (RG) sense
in $3+1$ dimensions, and is known \cite{peskin1978mandelstam} to
destabilize the continuous transition and render it weakly first
order.  Disorder alone is strongly relevant, and the transition
becomes non-trivial: a double epsilon expansion \cite{boyanovsky1982critical} exists for the critical theory, but  the extrapolation to $3+1$ dimensions is quantitatively poor.   Nevertheless, it supports a picture of a statistically scale-invariant theory, characterized by a dynamical exponent $z$ relating time and space, $t\sim x^z$ or frequency and wavevector, $\omega \sim k^z$, with $z>1$, reflecting the slow-down of dynamics by disorder.  

Now we consider the two effects together.  For very weak disorder the first order transition of the abelian Higgs theory is stable according to Imry-Ma arguments,\cite{imry1975} but it should be rapidly removed with stronger disorder.  To access the resulting continuous transition, we perturb the disordered critical point, which has some non-trivial critical action $S_d$, by coupling to the gauge field, and show that this coupling is relevant in the RG sense.  To do so, we write the action as $S=S_d + \int d^3{\bm x} d\tau\, [ i e A_\mu J_\mu+ F_{\mu\nu}^2 ]$, where $F_{\mu\nu}$ is the field strength of the emergent gauge field $A_\mu$, and $J^\mu({\bm x},\tau)$ are the $U(1)$ space-time currents of the bosons.  We adopt the Coulomb gauge ${\bm\nabla}\cdot {\bm A}=0$ and integrate out the gauge fields to obtain an effective long-range interaction between currents.  In particular, the time-components interact according to
 \begin{equation}
  \label{eq:12}
  S_{00} \sim e^2 \int\! d^3{\bm x} d^3{\bm x}' d\tau \frac{J^0({\bm x},\tau) J^0({\bm x}',\tau) }{|{\bm x}-{\bm x}'|}.
\end{equation}
Now we can proceed by using the fact that scaling dimensions of
conserved currents are unrenormalized, which has been demonstrated
even for disordered field theories \cite{wen1992scaling}.  This allows us to {\em exactly} power count Eq.~(\ref{eq:12}), according to $J^0 \sim L^{-d}$, and $\tau \sim L^z$.  We obtain $S_{00} \sim e^2 L^{z-1}$, which implies that the coupling $e^2$ is relevant for $z>1$.  Thus we predict the system flows to a new critical theory with both non-zero disorder and gauge coupling. This is a new quantum critical universality class not heretofore studied to our knowledge, the characterization of which is beyond the scope of this work but an interesting problem for future study.

{\em \underline{\textbf{Experiments and beyond.---}}} Probably the
most dramatic implication of our theory is that the well-studied and
characterized classical spin ice Ho$_2$Ti$_2$O$_7$ can be converted to
a QSL by introducing disorder.  Interestingly, the dynamics are
non-monotonic with disorder: introducing weak disorder first {\em
  speeds up} the dynamics by introducing transverse processes, while
strong disorder fully quenches and freezes the moments.  This
non-monotonicity should be visible in many observables, and notably
spin thermal conductivity, which has oft been considered a key
measurement in QSLs \cite{yamashita2010highly,yamashita2009thermal}.
Unfortunately its interpretation is typically clouded by the
difficulty of separating the (sought after) contribution from the
intrinsic heat conduction of the spins, from the (less interesting)
heat carried by phonons but {\em scattered} by spins.  Here the
non-monotonicity should aid in a clean separation of these effects: on
introducing disorder the spin thermal conductivity grows, developing a
large $T^3$ contribution, whose coefficient first increases, reaches a
maximum, and then collapses on leaving the QSL state.  Indications of
the disorder-catalyzed dynamics should be visible also in many other
probes, such as a NMR and NMQ relaxation, $\mu$SR, and microwave
conductivity.  Within the QSL state, the photon mode could be observed
in inelastic neutron scattering, with an intrinsic width controlled by
disorder, and growing with frequency.  In the Griffiths QSL, the
gapless localized electric excitations can also be pairwise excited,
introducing a momentum-independent background, which we expect scales
as $S(k,\omega)\sim e^{-c/\omega^x}$, with $x$ of order one.  A whole
range of other measurements should be possible to study scaling
properties at the quantum critical point terminating the QSL phase.
Our results may also be applicable to Pr$_2$Zr$_2$O$_7$, in which
random crystal field splittings have already been observed
\cite{kimura2013quantum}.   A slowly varying texture of the random
fields $h_i$, implicated there \cite{broholm2015}, does not reduce the stabilization of the QSL, which, as discussed above, even occurs for constant, non-random $h_i=h$.  

We finish by pointing out some connections of theoretical interest to
the active field of {\em many body localization} (MBL), which
describes systems in which ergodicity is violated at non-zero
temperature and eigenstates retain area law entanglement at non-zero
energy density \cite{nandkishore2015}.  In the strong random field
limit, our model falls into the MBL class, and we expect the trivial
phase displays those characteristics. For example, it has zero thermal
conductivity at any temperature. However, we guess that MBL occurs
only deep in the trivial regime, and that close to the QSL phases the
trivial state is insulating only at $T=0$.   In reality, the spins are
coupled to acoustic phonons which are always delocalized, preventing
MBL when this coupling is included.  However, it is interesting to
note that in the QSL phases, the delocalization of the photon plays a
similar role, and we conclude that, even in a closed system, MBL is
impossible for a Coulombic QSL. 

{\em Acknowledgements.---}This work was carried out while at KITP where L.S.\ was supported by Grants NSF-DMR-11-21053 and NSF-PHY-11-25915, and a postdoctoral fellowship from the Gordon and Betty Moore Foundation, under the EPiQS initiative, Grant GBMF-4303. L.B.\ was supported by DOE through BES grant number DE-FG02-08ER46524.

\nocite{chern2010disorder}

\bibliography{icing.bib}

\begin{thebibliography}{33}%
\makeatletter
\providecommand \@ifxundefined [1]{%
 \@ifx{#1\undefined}
}%
\providecommand \@ifnum [1]{%
 \ifnum #1\expandafter \@firstoftwo
 \else \expandafter \@secondoftwo
 \fi
}%
\providecommand \@ifx [1]{%
 \ifx #1\expandafter \@firstoftwo
 \else \expandafter \@secondoftwo
 \fi
}%
\providecommand \natexlab [1]{#1}%
\providecommand \enquote  [1]{``#1''}%
\providecommand \bibnamefont  [1]{#1}%
\providecommand \bibfnamefont [1]{#1}%
\providecommand \citenamefont [1]{#1}%
\providecommand \href@noop [0]{\@secondoftwo}%
\providecommand \href [0]{\begingroup \@sanitize@url \@href}%
\providecommand \@href[1]{\@@startlink{#1}\@@href}%
\providecommand \@@href[1]{\endgroup#1\@@endlink}%
\providecommand \@sanitize@url [0]{\catcode `\\12\catcode `\$12\catcode
  `\&12\catcode `\#12\catcode `\^12\catcode `\_12\catcode `\%12\relax}%
\providecommand \@@startlink[1]{}%
\providecommand \@@endlink[0]{}%
\providecommand \url  [0]{\begingroup\@sanitize@url \@url }%
\providecommand \@url [1]{\endgroup\@href {#1}{\urlprefix }}%
\providecommand \urlprefix  [0]{URL }%
\providecommand \Eprint [0]{\href }%
\providecommand \doibase [0]{http://dx.doi.org/}%
\providecommand \selectlanguage [0]{\@gobble}%
\providecommand \bibinfo  [0]{\@secondoftwo}%
\providecommand \bibfield  [0]{\@secondoftwo}%
\providecommand \translation [1]{[#1]}%
\providecommand \BibitemOpen [0]{}%
\providecommand \bibitemStop [0]{}%
\providecommand \bibitemNoStop [0]{.\EOS\space}%
\providecommand \EOS [0]{\spacefactor3000\relax}%
\providecommand \BibitemShut  [1]{\csname bibitem#1\endcsname}%
\let\auto@bib@innerbib\@empty
\bibitem [{\citenamefont {Amico}\ \emph {et~al.}(2008)\citenamefont {Amico},
  \citenamefont {Fazio}, \citenamefont {Osterloh},\ and\ \citenamefont
  {Vedral}}]{amico2008entanglement}%
  \BibitemOpen
  \bibfield  {author} {\bibinfo {author} {\bibfnamefont {L.}~\bibnamefont
  {Amico}}, \bibinfo {author} {\bibfnamefont {R.}~\bibnamefont {Fazio}},
  \bibinfo {author} {\bibfnamefont {A.}~\bibnamefont {Osterloh}}, \ and\
  \bibinfo {author} {\bibfnamefont {V.}~\bibnamefont {Vedral}},\ }\href
  {\doibase 10.1103/RevModPhys.80.517} {\bibfield  {journal} {\bibinfo
  {journal} {Rev. Mod. Phys.}\ }\textbf {\bibinfo {volume} {80}},\ \bibinfo
  {pages} {517} (\bibinfo {year} {2008})}\BibitemShut {NoStop}%
\bibitem [{\citenamefont {Chen}\ \emph {et~al.}(2010)\citenamefont {Chen},
  \citenamefont {Gu},\ and\ \citenamefont {Wen}}]{chen2010local}%
  \BibitemOpen
  \bibfield  {author} {\bibinfo {author} {\bibfnamefont {X.}~\bibnamefont
  {Chen}}, \bibinfo {author} {\bibfnamefont {Z.-C.}\ \bibnamefont {Gu}}, \ and\
  \bibinfo {author} {\bibfnamefont {X.-G.}\ \bibnamefont {Wen}},\ }\href
  {\doibase 10.1103/PhysRevB.82.155138} {\bibfield  {journal} {\bibinfo
  {journal} {Phys. Rev. B}\ }\textbf {\bibinfo {volume} {82}},\ \bibinfo
  {pages} {155138} (\bibinfo {year} {2010})}\BibitemShut {NoStop}%
\bibitem [{\citenamefont {Swingle}\ and\ \citenamefont
  {Van~Raamsdonk}(2014)}]{swingle2014universality}%
  \BibitemOpen
  \bibfield  {author} {\bibinfo {author} {\bibfnamefont {B.}~\bibnamefont
  {Swingle}}\ and\ \bibinfo {author} {\bibfnamefont {M.}~\bibnamefont
  {Van~Raamsdonk}},\ }\href {http://arxiv.org/abs/1405.2933} {\bibfield
  {journal} {\bibinfo  {journal} {arXiv preprint arXiv:1405.2933}\ } (\bibinfo
  {year} {2014})}\BibitemShut {NoStop}%
\bibitem [{\citenamefont {Balents}(2010)}]{balents2010spin}%
  \BibitemOpen
  \bibfield  {author} {\bibinfo {author} {\bibfnamefont {L.}~\bibnamefont
  {Balents}},\ }\href {\doibase 10.1038/nature08917} {\bibfield  {journal}
  {\bibinfo  {journal} {Nature}\ }\textbf {\bibinfo {volume} {464}},\ \bibinfo
  {pages} {199} (\bibinfo {year} {2010})}\BibitemShut {NoStop}%
\bibitem [{\citenamefont {Gingras}(2011)}]{gingras11:_spin_ice}%
  \BibitemOpen
  \bibfield  {author} {\bibinfo {author} {\bibfnamefont {M.}~\bibnamefont
  {Gingras}},\ }in\ \href@noop {} {\emph {\bibinfo {booktitle} {Introduction to
  Frustrated Magnetism}}},\ Vol.\ \bibinfo {volume} {164},\ \bibinfo {editor}
  {edited by\ \bibinfo {editor} {\bibfnamefont {C.}~\bibnamefont {Lacroix}},
  \bibinfo {editor} {\bibfnamefont {P.}~\bibnamefont {Mendels}}, \ and\
  \bibinfo {editor} {\bibfnamefont {F.}~\bibnamefont {Mila}}}\ (\bibinfo
  {publisher} {Springer Berlin Heidelberg},\ \bibinfo {year}
  {2011})\BibitemShut {NoStop}%
\bibitem [{\citenamefont {Altman}\ \emph {et~al.}(2004)\citenamefont {Altman},
  \citenamefont {Kafri}, \citenamefont {Polkovnikov},\ and\ \citenamefont
  {Refael}}]{altman2004phase}%
  \BibitemOpen
  \bibfield  {author} {\bibinfo {author} {\bibfnamefont {E.}~\bibnamefont
  {Altman}}, \bibinfo {author} {\bibfnamefont {Y.}~\bibnamefont {Kafri}},
  \bibinfo {author} {\bibfnamefont {A.}~\bibnamefont {Polkovnikov}}, \ and\
  \bibinfo {author} {\bibfnamefont {G.}~\bibnamefont {Refael}},\ }\href
  {\doibase 10.1103/PhysRevLett.93.150402} {\bibfield  {journal} {\bibinfo
  {journal} {Phys. Rev. Lett.}\ }\textbf {\bibinfo {volume} {93}},\ \bibinfo
  {pages} {150402} (\bibinfo {year} {2004})}\BibitemShut {NoStop}%
\bibitem [{\citenamefont {Weichman}(2008)}]{weichman2008dirty}%
  \BibitemOpen
  \bibfield  {author} {\bibinfo {author} {\bibfnamefont {P.~B.}\ \bibnamefont
  {Weichman}},\ }\href {\doibase 10.1142/S0217984908017187} {\bibfield
  {journal} {\bibinfo  {journal} {Modern Physics Letters B}\ }\textbf {\bibinfo
  {volume} {22}},\ \bibinfo {pages} {2623} (\bibinfo {year}
  {2008})}\BibitemShut {NoStop}%
\bibitem [{\citenamefont {Fisher}\ \emph {et~al.}(1989)\citenamefont {Fisher},
  \citenamefont {Weichman}, \citenamefont {Grinstein},\ and\ \citenamefont
  {Fisher}}]{fisher1989boson}%
  \BibitemOpen
  \bibfield  {author} {\bibinfo {author} {\bibfnamefont {M.~P.}\ \bibnamefont
  {Fisher}}, \bibinfo {author} {\bibfnamefont {P.~B.}\ \bibnamefont
  {Weichman}}, \bibinfo {author} {\bibfnamefont {G.}~\bibnamefont {Grinstein}},
  \ and\ \bibinfo {author} {\bibfnamefont {D.~S.}\ \bibnamefont {Fisher}},\
  }\href {\doibase 10.1103/PhysRevB.40.546} {\bibfield  {journal} {\bibinfo
  {journal} {Phys. Rev. B}\ }\textbf {\bibinfo {volume} {40}},\ \bibinfo
  {pages} {546} (\bibinfo {year} {1989})}\BibitemShut {NoStop}%
\bibitem [{\citenamefont {Gardner}\ \emph {et~al.}(2010)\citenamefont
  {Gardner}, \citenamefont {Gingras},\ and\ \citenamefont
  {Greedan}}]{gardner2010magnetic}%
  \BibitemOpen
  \bibfield  {author} {\bibinfo {author} {\bibfnamefont {J.~S.}\ \bibnamefont
  {Gardner}}, \bibinfo {author} {\bibfnamefont {M.~J.}\ \bibnamefont
  {Gingras}}, \ and\ \bibinfo {author} {\bibfnamefont {J.~E.}\ \bibnamefont
  {Greedan}},\ }\href {\doibase 10.1103/RevModPhys.82.53} {\bibfield  {journal}
  {\bibinfo  {journal} {Rev. Mod. Phys.}\ }\textbf {\bibinfo {volume} {82}},\
  \bibinfo {pages} {53} (\bibinfo {year} {2010})}\BibitemShut {NoStop}%
\bibitem [{\citenamefont {Isakov}\ \emph {et~al.}(2005)\citenamefont {Isakov},
  \citenamefont {Moessner},\ and\ \citenamefont {Sondhi}}]{isakov2005spin}%
  \BibitemOpen
  \bibfield  {author} {\bibinfo {author} {\bibfnamefont {S.~V.}\ \bibnamefont
  {Isakov}}, \bibinfo {author} {\bibfnamefont {R.}~\bibnamefont {Moessner}}, \
  and\ \bibinfo {author} {\bibfnamefont {S.}~\bibnamefont {Sondhi}},\ }\href
  {\doibase 10.1103/PhysRevLett.95.217201} {\bibfield  {journal} {\bibinfo
  {journal} {Phys. Rev. Lett.}\ }\textbf {\bibinfo {volume} {95}},\ \bibinfo
  {pages} {217201} (\bibinfo {year} {2005})}\BibitemShut {NoStop}%
\bibitem [{\citenamefont {Kimura}\ \emph {et~al.}(2013)\citenamefont {Kimura},
  \citenamefont {Nakatsuji}, \citenamefont {Wen}, \citenamefont {Broholm},
  \citenamefont {Stone}, \citenamefont {Nishibori},\ and\ \citenamefont
  {Sawa}}]{kimura2013quantum}%
  \BibitemOpen
  \bibfield  {author} {\bibinfo {author} {\bibfnamefont {K.}~\bibnamefont
  {Kimura}}, \bibinfo {author} {\bibfnamefont {S.}~\bibnamefont {Nakatsuji}},
  \bibinfo {author} {\bibfnamefont {J.}~\bibnamefont {Wen}}, \bibinfo {author}
  {\bibfnamefont {C.}~\bibnamefont {Broholm}}, \bibinfo {author} {\bibfnamefont
  {M.}~\bibnamefont {Stone}}, \bibinfo {author} {\bibfnamefont
  {E.}~\bibnamefont {Nishibori}}, \ and\ \bibinfo {author} {\bibfnamefont
  {H.}~\bibnamefont {Sawa}},\ }\href {\doibase 10.1038/ncomms2914} {\bibfield
  {journal} {\bibinfo  {journal} {Nat. Commun.}\ }\textbf {\bibinfo {volume}
  {4}} (\bibinfo {year} {2013}),\ 10.1038/ncomms2914}\BibitemShut {NoStop}%
\bibitem [{\citenamefont {Hermele}\ \emph {et~al.}(2004)\citenamefont
  {Hermele}, \citenamefont {Fisher},\ and\ \citenamefont
  {Balents}}]{hermele2004pyrochlore}%
  \BibitemOpen
  \bibfield  {author} {\bibinfo {author} {\bibfnamefont {M.}~\bibnamefont
  {Hermele}}, \bibinfo {author} {\bibfnamefont {M.~P.}\ \bibnamefont {Fisher}},
  \ and\ \bibinfo {author} {\bibfnamefont {L.}~\bibnamefont {Balents}},\ }\href
  {\doibase 10.1103/PhysRevB.69.064404} {\bibfield  {journal} {\bibinfo
  {journal} {Phys. Rev. B}\ }\textbf {\bibinfo {volume} {69}},\ \bibinfo
  {pages} {064404} (\bibinfo {year} {2004})}\BibitemShut {NoStop}%
\bibitem [{\citenamefont {Banks}\ \emph {et~al.}(1977)\citenamefont {Banks},
  \citenamefont {Myerson},\ and\ \citenamefont {Kogut}}]{banks1977phase}%
  \BibitemOpen
  \bibfield  {author} {\bibinfo {author} {\bibfnamefont {T.}~\bibnamefont
  {Banks}}, \bibinfo {author} {\bibfnamefont {R.}~\bibnamefont {Myerson}}, \
  and\ \bibinfo {author} {\bibfnamefont {J.}~\bibnamefont {Kogut}},\ }\href
  {\doibase 10.1016/0550-3213(77)90129-8} {\bibfield  {journal} {\bibinfo
  {journal} {Nuclear Physics B}\ }\textbf {\bibinfo {volume} {129}},\ \bibinfo
  {pages} {493} (\bibinfo {year} {1977})}\BibitemShut {NoStop}%
\bibitem [{\citenamefont {Banerjee}\ \emph {et~al.}(2008)\citenamefont
  {Banerjee}, \citenamefont {Isakov}, \citenamefont {Damle},\ and\
  \citenamefont {Kim}}]{banerjee2008unusual}%
  \BibitemOpen
  \bibfield  {author} {\bibinfo {author} {\bibfnamefont {A.}~\bibnamefont
  {Banerjee}}, \bibinfo {author} {\bibfnamefont {S.~V.}\ \bibnamefont
  {Isakov}}, \bibinfo {author} {\bibfnamefont {K.}~\bibnamefont {Damle}}, \
  and\ \bibinfo {author} {\bibfnamefont {Y.~B.}\ \bibnamefont {Kim}},\ }\href
  {\doibase 10.1103/PhysRevLett.100.047208} {\bibfield  {journal} {\bibinfo
  {journal} {Phys. Rev. Lett.}\ }\textbf {\bibinfo {volume} {100}},\ \bibinfo
  {pages} {047208} (\bibinfo {year} {2008})}\BibitemShut {NoStop}%
\bibitem [{\citenamefont {Shannon}\ \emph {et~al.}(2012)\citenamefont
  {Shannon}, \citenamefont {Sikora}, \citenamefont {Pollmann}, \citenamefont
  {Penc},\ and\ \citenamefont {Fulde}}]{shannon2012quantum}%
  \BibitemOpen
  \bibfield  {author} {\bibinfo {author} {\bibfnamefont {N.}~\bibnamefont
  {Shannon}}, \bibinfo {author} {\bibfnamefont {O.}~\bibnamefont {Sikora}},
  \bibinfo {author} {\bibfnamefont {F.}~\bibnamefont {Pollmann}}, \bibinfo
  {author} {\bibfnamefont {K.}~\bibnamefont {Penc}}, \ and\ \bibinfo {author}
  {\bibfnamefont {P.}~\bibnamefont {Fulde}},\ }\href {\doibase
  10.1103/PhysRevLett.108.067204} {\bibfield  {journal} {\bibinfo  {journal}
  {Phys. Rev. Lett.}\ }\textbf {\bibinfo {volume} {108}},\ \bibinfo {pages}
  {067204} (\bibinfo {year} {2012})}\BibitemShut {NoStop}%
\bibitem [{\citenamefont {Kato}\ and\ \citenamefont {Onoda}(2015)}]{kato2015}%
  \BibitemOpen
  \bibfield  {author} {\bibinfo {author} {\bibfnamefont {Y.}~\bibnamefont
  {Kato}}\ and\ \bibinfo {author} {\bibfnamefont {S.}~\bibnamefont {Onoda}},\
  }\href {\doibase 10.1103/PhysRevLett.115.077202} {\bibfield  {journal}
  {\bibinfo  {journal} {Phys. Rev. Lett.}\ }\textbf {\bibinfo {volume} {115}},\
  \bibinfo {pages} {077202} (\bibinfo {year} {2015})}\BibitemShut {NoStop}%
\bibitem [{\citenamefont {John}\ \emph {et~al.}(1983)\citenamefont {John},
  \citenamefont {Sompolinsky},\ and\ \citenamefont
  {Stephen}}]{PhysRevB.27.5592}%
  \BibitemOpen
  \bibfield  {author} {\bibinfo {author} {\bibfnamefont {S.}~\bibnamefont
  {John}}, \bibinfo {author} {\bibfnamefont {H.}~\bibnamefont {Sompolinsky}}, \
  and\ \bibinfo {author} {\bibfnamefont {M.~J.}\ \bibnamefont {Stephen}},\
  }\href {\doibase 10.1103/PhysRevB.27.5592} {\bibfield  {journal} {\bibinfo
  {journal} {Phys. Rev. B}\ }\textbf {\bibinfo {volume} {27}},\ \bibinfo
  {pages} {5592} (\bibinfo {year} {1983})}\BibitemShut {NoStop}%
\bibitem [{\citenamefont {John}(1991)}]{john1991localization}%
  \BibitemOpen
  \bibfield  {author} {\bibinfo {author} {\bibfnamefont {S.}~\bibnamefont
  {John}},\ }\href {\doibase 10.1063/1.881300} {\bibfield  {journal} {\bibinfo
  {journal} {Phys. Today}\ }\textbf {\bibinfo {volume} {44}},\ \bibinfo {pages}
  {32} (\bibinfo {year} {1991})}\BibitemShut {NoStop}%
\bibitem [{\citenamefont {Villain}\ \emph {et~al.}(1980)\citenamefont
  {Villain}, \citenamefont {Bidaux}, \citenamefont {Carton},\ and\
  \citenamefont {Conte}}]{villain1980order}%
  \BibitemOpen
  \bibfield  {author} {\bibinfo {author} {\bibfnamefont {J.}~\bibnamefont
  {Villain}}, \bibinfo {author} {\bibfnamefont {R.}~\bibnamefont {Bidaux}},
  \bibinfo {author} {\bibfnamefont {J.-P.}\ \bibnamefont {Carton}}, \ and\
  \bibinfo {author} {\bibfnamefont {R.}~\bibnamefont {Conte}},\ }\href
  {\doibase 10.1051/jphys: 0198000410110126300} {\bibfield  {journal} {\bibinfo
   {journal} {Journal de Physique}\ }\textbf {\bibinfo {volume} {41}},\
  \bibinfo {pages} {1263} (\bibinfo {year} {1980})}\BibitemShut {NoStop}%
\bibitem [{\citenamefont {Henley}(1989)}]{henley1989ordering}%
  \BibitemOpen
  \bibfield  {author} {\bibinfo {author} {\bibfnamefont {C.~L.}\ \bibnamefont
  {Henley}},\ }\href {\doibase 10.1103/PhysRevLett.62.2056} {\bibfield
  {journal} {\bibinfo  {journal} {Phys. Rev. Lett.}\ }\textbf {\bibinfo
  {volume} {62}},\ \bibinfo {pages} {2056} (\bibinfo {year}
  {1989})}\BibitemShut {NoStop}%
\bibitem [{\citenamefont {Savary}\ and\ \citenamefont
  {Balents}(2012)}]{savary2012coulombic}%
  \BibitemOpen
  \bibfield  {author} {\bibinfo {author} {\bibfnamefont {L.}~\bibnamefont
  {Savary}}\ and\ \bibinfo {author} {\bibfnamefont {L.}~\bibnamefont
  {Balents}},\ }\href {\doibase 10.1103/PhysRevLett.108.037202} {\bibfield
  {journal} {\bibinfo  {journal} {Phys. Rev. Lett.}\ }\textbf {\bibinfo
  {volume} {108}},\ \bibinfo {pages} {037202} (\bibinfo {year}
  {2012})}\BibitemShut {NoStop}%
\bibitem [{\citenamefont {Griffiths}(1969)}]{griffiths1969}%
  \BibitemOpen
  \bibfield  {author} {\bibinfo {author} {\bibfnamefont {R.~B.}\ \bibnamefont
  {Griffiths}},\ }\href {\doibase 10.1103/PhysRevLett.23.17} {\bibfield
  {journal} {\bibinfo  {journal} {Phys. Rev. Lett.}\ }\textbf {\bibinfo
  {volume} {23}},\ \bibinfo {pages} {17} (\bibinfo {year} {1969})}\BibitemShut
  {NoStop}%
\bibitem [{\citenamefont {Iyer}\ \emph {et~al.}(2012)\citenamefont {Iyer},
  \citenamefont {Pekker},\ and\ \citenamefont {Refael}}]{iyer2012mott}%
  \BibitemOpen
  \bibfield  {author} {\bibinfo {author} {\bibfnamefont {S.}~\bibnamefont
  {Iyer}}, \bibinfo {author} {\bibfnamefont {D.}~\bibnamefont {Pekker}}, \ and\
  \bibinfo {author} {\bibfnamefont {G.}~\bibnamefont {Refael}},\ }\href
  {\doibase 10.1103/PhysRevB.85.094202} {\bibfield  {journal} {\bibinfo
  {journal} {Phys. Rev. B}\ }\textbf {\bibinfo {volume} {85}},\ \bibinfo
  {pages} {094202} (\bibinfo {year} {2012})}\BibitemShut {NoStop}%
\bibitem [{\citenamefont {Deutscher}(2006)}]{deutscher2006}%
  \BibitemOpen
  \bibfield  {author} {\bibinfo {author} {\bibfnamefont {G.}~\bibnamefont
  {Deutscher}},\ }\href@noop {} {\emph {\bibinfo {title} {New Superconductors:
  From Granular to High {Tc}}}}\ (\bibinfo  {publisher} {World Scientific
  Publishing Co.},\ \bibinfo {year} {2006})\BibitemShut {NoStop}%
\bibitem [{\citenamefont {Peskin}(1978)}]{peskin1978mandelstam}%
  \BibitemOpen
  \bibfield  {author} {\bibinfo {author} {\bibfnamefont {M.~E.}\ \bibnamefont
  {Peskin}},\ }\href {\doibase 10.1016/0003-4916(78)90252-X} {\bibfield
  {journal} {\bibinfo  {journal} {Annals of Physics}\ }\textbf {\bibinfo
  {volume} {113}},\ \bibinfo {pages} {122} (\bibinfo {year}
  {1978})}\BibitemShut {NoStop}%
\bibitem [{\citenamefont {Boyanovsky}\ and\ \citenamefont
  {Cardy}(1982)}]{boyanovsky1982critical}%
  \BibitemOpen
  \bibfield  {author} {\bibinfo {author} {\bibfnamefont {D.}~\bibnamefont
  {Boyanovsky}}\ and\ \bibinfo {author} {\bibfnamefont {J.~L.}\ \bibnamefont
  {Cardy}},\ }\href {\doibase 10.1103/PhysRevB.26.154} {\bibfield  {journal}
  {\bibinfo  {journal} {Phys. Rev. B}\ }\textbf {\bibinfo {volume} {26}},\
  \bibinfo {pages} {154} (\bibinfo {year} {1982})}\BibitemShut {NoStop}%
\bibitem [{\citenamefont {Imry}\ and\ \citenamefont {Ma}(1975)}]{imry1975}%
  \BibitemOpen
  \bibfield  {author} {\bibinfo {author} {\bibfnamefont {Y.}~\bibnamefont
  {Imry}}\ and\ \bibinfo {author} {\bibfnamefont {S.-k.}\ \bibnamefont {Ma}},\
  }\href {\doibase 10.1103/PhysRevLett.35.1399} {\bibfield  {journal} {\bibinfo
   {journal} {Phys. Rev. Lett.}\ }\textbf {\bibinfo {volume} {35}},\ \bibinfo
  {pages} {1399} (\bibinfo {year} {1975})}\BibitemShut {NoStop}%
\bibitem [{\citenamefont {Wen}(1992)}]{wen1992scaling}%
  \BibitemOpen
  \bibfield  {author} {\bibinfo {author} {\bibfnamefont {X.-G.}\ \bibnamefont
  {Wen}},\ }\href {\doibase 10.1103/PhysRevB.46.2655} {\bibfield  {journal}
  {\bibinfo  {journal} {Phys. Rev. B}\ }\textbf {\bibinfo {volume} {46}},\
  \bibinfo {pages} {2655} (\bibinfo {year} {1992})}\BibitemShut {NoStop}%
\bibitem [{\citenamefont {Yamashita}\ \emph {et~al.}(2010)\citenamefont
  {Yamashita}, \citenamefont {Nakata}, \citenamefont {Senshu}, \citenamefont
  {Nagata}, \citenamefont {Yamamoto}, \citenamefont {Kato}, \citenamefont
  {Shibauchi},\ and\ \citenamefont {Matsuda}}]{yamashita2010highly}%
  \BibitemOpen
  \bibfield  {author} {\bibinfo {author} {\bibfnamefont {M.}~\bibnamefont
  {Yamashita}}, \bibinfo {author} {\bibfnamefont {N.}~\bibnamefont {Nakata}},
  \bibinfo {author} {\bibfnamefont {Y.}~\bibnamefont {Senshu}}, \bibinfo
  {author} {\bibfnamefont {M.}~\bibnamefont {Nagata}}, \bibinfo {author}
  {\bibfnamefont {H.~M.}\ \bibnamefont {Yamamoto}}, \bibinfo {author}
  {\bibfnamefont {R.}~\bibnamefont {Kato}}, \bibinfo {author} {\bibfnamefont
  {T.}~\bibnamefont {Shibauchi}}, \ and\ \bibinfo {author} {\bibfnamefont
  {Y.}~\bibnamefont {Matsuda}},\ }\href {\doibase 10.1126/science.1188200}
  {\bibfield  {journal} {\bibinfo  {journal} {Science}\ }\textbf {\bibinfo
  {volume} {328}},\ \bibinfo {pages} {1246} (\bibinfo {year}
  {2010})}\BibitemShut {NoStop}%
\bibitem [{\citenamefont {Yamashita}\ \emph {et~al.}(2009)\citenamefont
  {Yamashita}, \citenamefont {Nakata}, \citenamefont {Kasahara}, \citenamefont
  {Sasaki}, \citenamefont {Yoneyama}, \citenamefont {Kobayashi}, \citenamefont
  {Fujimoto}, \citenamefont {Shibauchi},\ and\ \citenamefont
  {Matsuda}}]{yamashita2009thermal}%
  \BibitemOpen
  \bibfield  {author} {\bibinfo {author} {\bibfnamefont {M.}~\bibnamefont
  {Yamashita}}, \bibinfo {author} {\bibfnamefont {N.}~\bibnamefont {Nakata}},
  \bibinfo {author} {\bibfnamefont {Y.}~\bibnamefont {Kasahara}}, \bibinfo
  {author} {\bibfnamefont {T.}~\bibnamefont {Sasaki}}, \bibinfo {author}
  {\bibfnamefont {N.}~\bibnamefont {Yoneyama}}, \bibinfo {author}
  {\bibfnamefont {N.}~\bibnamefont {Kobayashi}}, \bibinfo {author}
  {\bibfnamefont {S.}~\bibnamefont {Fujimoto}}, \bibinfo {author}
  {\bibfnamefont {T.}~\bibnamefont {Shibauchi}}, \ and\ \bibinfo {author}
  {\bibfnamefont {Y.}~\bibnamefont {Matsuda}},\ }\href {\doibase
  10.1038/nphys1134} {\bibfield  {journal} {\bibinfo  {journal} {Nat. Phys.}\
  }\textbf {\bibinfo {volume} {5}},\ \bibinfo {pages} {44} (\bibinfo {year}
  {2009})}\BibitemShut {NoStop}%
\bibitem [{\citenamefont {Broholm}(2015)}]{broholm2015}%
  \BibitemOpen
  \bibfield  {author} {\bibinfo {author} {\bibfnamefont {C.}~\bibnamefont
  {Broholm}},\ }\href@noop {} {\bibfield  {journal} {\bibinfo  {journal}
  {private communication}\ } (\bibinfo {year} {2015})}\BibitemShut {NoStop}%
\bibitem [{\citenamefont {Nandkishore}\ and\ \citenamefont
  {Huse}(2015)}]{nandkishore2015}%
  \BibitemOpen
  \bibfield  {author} {\bibinfo {author} {\bibfnamefont {R.}~\bibnamefont
  {Nandkishore}}\ and\ \bibinfo {author} {\bibfnamefont {D.~A.}\ \bibnamefont
  {Huse}},\ }\href {\doibase 10.1146/annurev-conmatphys-031214-014726}
  {\bibfield  {journal} {\bibinfo  {journal} {Annual Review of Condensed Matter
  Physics}\ }\textbf {\bibinfo {volume} {6}},\ \bibinfo {pages} {15} (\bibinfo
  {year} {2015})}\BibitemShut {NoStop}%
\bibitem [{\citenamefont {Chern}\ \emph {et~al.}(2010)\citenamefont {Chern},
  \citenamefont {Liao},\ and\ \citenamefont {Chou}}]{chern2010disorder}%
  \BibitemOpen
  \bibfield  {author} {\bibinfo {author} {\bibfnamefont {C.-H.}\ \bibnamefont
  {Chern}}, \bibinfo {author} {\bibfnamefont {C.-N.}\ \bibnamefont {Liao}}, \
  and\ \bibinfo {author} {\bibfnamefont {Y.-Z.}\ \bibnamefont {Chou}},\
  }\href@noop {} {\bibfield  {journal} {\bibinfo  {journal} {arXiv preprint
  arXiv:1003.4204}\ } (\bibinfo {year} {2010})}\BibitemShut {NoStop}%
\end{thebibliography}%

\appendix
\section{Definitions}
\label{sec:definitions}

\subsection{Local bases}

The local cubic bases in which the Hamiltonian Eq.~(3) is expressed are the following $(\mathbf{\hat{a}}_i,\mathbf{\hat{b}}_i,\mathbf{\hat{e}}_i)$ bases 
\begin{equation}
\left\{\begin{array}{l}
\mathbf{\hat{e}}_0=(1,1,1)/\sqrt{3}\\
\mathbf{\hat{e}}_1=(1,-1,-1)/\sqrt{3}\\
\mathbf{\hat{e}}_2=(-1,1,-1)/\sqrt{3}\\
\mathbf{\hat{e}}_3=(-1,-1,1)/\sqrt{3},
\end{array}\right.,
\quad
\left\{\begin{array}{l}
\mathbf{\hat{a}}_0=(-2,1,1)/\sqrt{6}\\
\mathbf{\hat{a}}_1=(-2,-1,-1)/\sqrt{6}\\
\mathbf{\hat{a}}_2=(2,1,-1)/\sqrt{6}\\
\mathbf{\hat{a}}_3=(2,-1,1)/\sqrt{6}
\end{array}\right.,
\end{equation}
$\mathbf{\hat{b}}_i=\mathbf{\hat{e}}_i\times\mathbf{\hat{a}}_i$, such that spin $\mathbf{S}_i$ on sublattice $i$ is $\mathbf{S}_i=\mathsf{S}^+_i(\mathbf{\hat{a}}_i-i\mathbf{\hat{b}}_i)/2+\mathsf{S}^-_i(\mathbf{\hat{a}}_i+i\mathbf{\hat{b}}_i)/2+\mathsf{S}^z_i\mathbf{\hat{e}}_i$.

\subsection{Lattice vectors}

The four nearest-neighbor vectors of a A-sublattice diamond site (sublattice A corresponds to ``up'' tetrahedra) are $\mathbf{e}_\mu=\frac{a\sqrt{3}}{4}\mathbf{\hat{e}}_\mu$, where $a$ is the usual FCC lattice spacing. The four pyrochlore sites of the ``up'' tetrahedron centered at the origin are located at $\mathbf{e}_\mu/2$, $\mu=0,..,3$. 

The FCC primitive lattice vectors are $\mathbf{A}_i=\mathbf{e}_0-\mathbf{e}_i$, $i=1,..,3$, while the reciprocal lattice basis vectors are defined as usual by $\mathbf{B}_1=2\pi\frac{\mathbf{A}_2\times\mathbf{A}_3}{v_{{\rm u.c.}}}$ and its cyclic permutations, where $v_{{\rm u.c.}}=\mathbf{A}_1\cdot(\mathbf{A}_2\times\mathbf{A}_3)$ is the volume of the (real space) unit cell.  If the $q_i$'s are defined as 
\begin{equation}
\label{eq:qi}
\mathbf{k}=\sum_{i=1}^3 q_i\,\mathbf{B}_i,
\end{equation} 
the first Brillouin zone can be considered the ``cube'' with unit sides described by $-1/2<q_i<1/2$ (note that the $q_i$'s are dimensionless).

\section{Details of perturbation theory}
\label{sec:deta-pert-theory}

Here, $H'=-\sum_ih_i\mathsf{S}_i^x=-\frac{1}{2}\sum_ih_i(\mathsf{S}_i^-+\mathsf{S}_i^+)$, for $|h_i|\ll J$, is treated as a perturbation over the highly-degenerate ground state manifold of the spin ice Hamiltonian $H_0=J\sum_{\langle ij\rangle}\mathsf{S}_i^z\mathsf{S}_j^z$. The degenerate manifold is that of all ``two-in-two-out'' tetrahedral configurations of the $\mathsf{S}^z$ components of the spins, where, on each tetrahedron $t$, $\sum_{i\in t}\mathsf{S}_i^z=0$. Similarly to the case of the transverse exchange perturbative terms $\mathsf{S}^+_i\mathsf{S}_j^-$, the lowest-order effective Hamiltonian obtained within degenerate perturbation theory is a ``ring-exchange'' term. In the present case, it arises at sixth order in $h/J$. The combinatorial coefficient of the ring exchange term is obtained by choosing one of the six hexagon sites in $\sum_{i}$ at each step of the perturbation theory, with only the constraint that $\mathsf{S}^+$ and $\mathsf{S}^-$ should alternate along the hexagon ring:
\begin{eqnarray}
  \label{eq:9}
&&H_{\rm eff}=\frac{1}{2^6}\sum_{i_1}\cdots\sum_{i_6} h_{i_1}\cdots
h_{i_6}\\
&&\qquad\times \mathcal{P}\left[(\mathsf{S}_{i_1}^++\mathsf{S}_{i_1}^-)\frac{1-\mathcal{P}}{H_0-E_0}\cdots\frac{1-\mathcal{P}}{H_0-E_0}(\mathsf{S}_{i_6}^++\mathsf{S}_{i_6}^-)\right]\mathcal{P}\nonumber
\end{eqnarray}
where $\mathcal{P}$ is the projection operator onto the degenerate
``2-in-2-out'' manifold and $E_0$ is the energy of any (static)
2-in-2-out configuration. The explicit calculation involves
considering which site is selected in the sum at each order of
perturbation theory. Different configurations with different energies
can occur, and one must keep track of all those coefficients. A
counting procedure is pictured in Fig.~3. 
The final coefficient of $63/16$ given in the main text (also given in Ref.~\onlinecite{chern2010disorder}) is obtained by summing over all the paths (including the energy denominator at each step). For example one may group (factor) the paths which go through the same configuration in the middle column of Fig.~3:
\begin{eqnarray}
  \label{eq:15}
&&\frac{1}{2^6}\frac{6}{J}\left\{\left(\frac{2}{J}+\frac{1}{2J}\right)\left(\frac{2}{J}\frac{2}{J}+\frac{2}{2J}\frac{1}{J}\right)\right.\\
&&\qquad\qquad+\left(\frac{1}{J}+\frac{2}{2J}\right)\left(\frac{2}{J}\frac{2}{2J}+\frac{1}{2J}\frac{4}{2J}+\frac{2}{2J}\frac{2}{2J}\right)\nonumber\\
&&\left.\qquad\qquad+\frac{3}{2J}\frac{2}{2J}\frac{1}{3J}\right\}\frac{2}{J}\nonumber\\
&&=\frac{63}{16J^5}.\nonumber
\end{eqnarray}

\begin{figure}
\includegraphics[width=\linewidth]{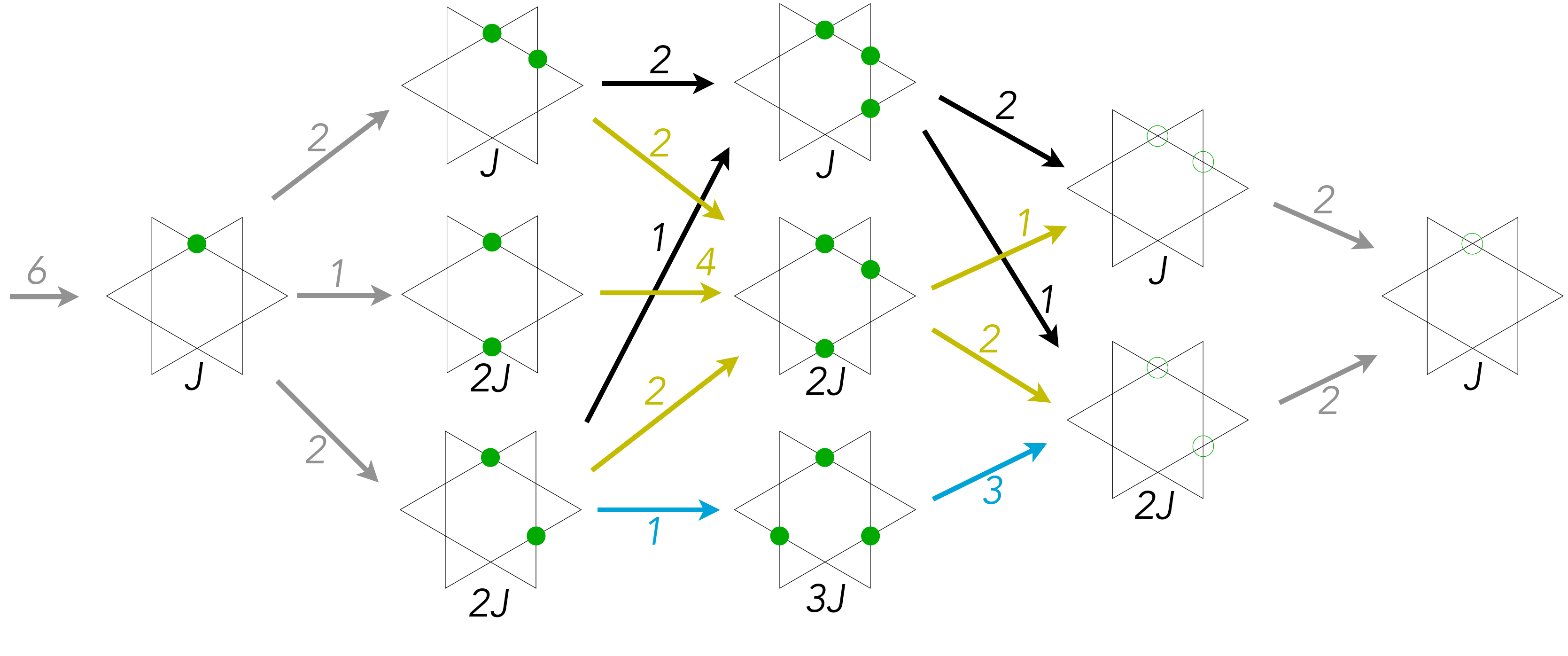}
  \caption{Perturbation theory paths. The filled blue dots represent which kinds of sites have been acted on at each step, and the energy of the configurations at each step is given below the hexagon stars. The empty circles represents sites which have not yet been acted upon. The numbers on the arrows indicate how many ways one can go between the configuration on the left and that on the right.}
  \label{fig:PT}
\end{figure}

\section{Details of gauge Mean Field Theory}
\label{sec:details-gauge-mean}

Here, we proceed like in Ref.~\onlinecite{savary2012coulombic}. We consider the Hamiltonian Eq.~(3)
\begin{equation}
  H=J\sum_{\langle ij\rangle}\mathsf{S}_i^z\mathsf{S}_j^z-\frac{1}{2}\sum_{i}h_i\left(\mathsf{S}_i^++\mathsf{S}_i^-\right).
\end{equation}
We rewrite:
\begin{equation}
\label{eq:13}
\begin{cases}
\mathsf{S}_{\mathbf{r},\mathbf{r}+\mathbf{e}_\mu}^+=\Phi_\mathbf{r}^\dagger\mathsf{s}_{\mathbf{r},\mathbf{r}+\mathbf{e}_\mu}^+\Phi_{\mathbf{r}+\mathbf{e}_\mu}\\
\mathsf{S}^z_{\mathbf{r},\mathbf{r}+\mathbf{e}_\mu}=\mathsf{s}^z_{\mathbf{r},\mathbf{r}+\mathbf{e}_\mu}
\end{cases}
\quad\mbox{for }\mathbf{r}\in{\rm A},
\end{equation}
where $\mathbf{r}$ label tetrahedra centers (dual diamond lattice sites) ---in Eq.~(\ref{eq:13}) the ``up'' tetrahedra (A sublattice)--- and the $\mathbf{e}_\mu$ are defined in Appendix~\ref{sec:definitions} and
\begin{equation}
Q_\mathbf{r}=\epsilon_\mathbf{r}\sum_\mu \mathsf{S}^z_{\mathbf{r},\mathbf{r}+\epsilon_\mathbf{r}\mathbf{e}_\mu},\qquad\epsilon_{\mathbf{r}}=\pm1\;\mbox{for}\;\mathbf{r}\in{\rm A,B}
\end{equation}
as defined in the main text. So:
\begin{eqnarray}
  H&=&\frac{J}{2}\sum_{\mathbf{r}\in{\rm
      A,B}}Q_\mathbf{r}^2\\
&&-\frac{1}{2}\sum_{\mathbf{r}\in{\rm
      A}}\sum_\mu h_{\mathbf{r},\mathbf{r}+\mathbf{e}_\mu}\\
&&\qquad\qquad\quad\times\left(\mathsf{s}_{\mathbf{r},\mathbf{r}+\mathbf{e}_\mu}^+\Phi_\mathbf{r}^\dagger\Phi_{\mathbf{r}+\mathbf{e}_\mu}+\mathsf{s}_{\mathbf{r},\mathbf{r}+\mathbf{e}_\mu}^-\Phi_{\mathbf{r}+\mathbf{e}_\mu}^\dagger\Phi_{\mathbf{r}}\right)\nonumber\\
&=&\frac{J}{2}\sum_{\mathbf{r}\in{\rm
      A,B}}Q_\mathbf{r}^2\\
&&-\frac{1}{2}\sum_{\mathbf{r}\in{\rm
      A,B}}\sum_\mu h_{\mathbf{r},\mathbf{r}+\epsilon_\mathbf{r}\mathbf{e}_\mu}\left(\mathsf{s}_{\mathbf{r},\mathbf{r}+\epsilon_\mathbf{r}\mathbf{e}_\mu}^{\epsilon_\mathbf{r}}\Phi_\mathbf{r}^\dagger\Phi_{\mathbf{r}+\epsilon_\mathbf{r}\mathbf{e}_\mu}\right).
\end{eqnarray}
We now perform the mean field decoupling:
\begin{equation}
\mathsf{s}\,\Phi^\dagger\Phi\rightarrow\langle\mathsf{s}\rangle\Phi^\dagger\Phi+\mathsf{s}\langle\Phi^\dagger\Phi\rangle-\langle\mathsf{s}\rangle\langle\Phi^\dagger\Phi\rangle,
\end{equation}
which yields:
\begin{equation}
  H_{\sf u}=-\sum_{\mathbf{r}\in{\rm A}}\sum_\mu\vec{\mathsf{u}}_{\mathbf{r},\mathbf{r}+\mathbf{e}_\mu}\cdot\vec{\mathsf{s}}_{\mathbf{r},\mathbf{r}+\mathbf{e}_\mu},
\end{equation}
where
\begin{equation}
  \label{eq:11}
\begin{cases}
\mathsf{u}^-_{\mathbf{r},\mathbf{r}+\mathbf{e}_\mu}=h_{\mathbf{r},\mathbf{r}+\mathbf{e}_\mu}\langle\Phi_\mathbf{r}^\dagger\Phi_{\mathbf{r}+\mathbf{e}_\mu}\rangle,\\
\mathsf{u}^+_{\mathbf{r},\mathbf{r}+\mathbf{e}_\mu}=\left(\mathsf{u}^+_{\mathbf{r},\mathbf{r}+\mathbf{e}_\mu}\right)^*,\\
\mathsf{u}^z_{\mathbf{r},\mathbf{r}+\mathbf{e}_\mu}=0,
\end{cases}
\end{equation}
and
\begin{equation}
  H_\Phi=-\frac{1}{2}\sum_{\mathbf{r}\in{\rm A,B}}\sum_\mu h_{\mathbf{r},\mathbf{r}+\epsilon_\mathbf{r}\mathbf{e}_\mu}\langle\mathsf{s}_{\mathbf{r},\mathbf{r}+\epsilon_\mathbf{r}\mathbf{e}_\mu}^{\epsilon_\mathbf{r}}\rangle\Phi_\mathbf{r}^\dagger\Phi_{\mathbf{r}+\epsilon_\mathbf{r}\mathbf{e}_\mu}.
\end{equation}
We now need to solve this self-consistently, bearing in mind the
constraint $\Phi_\mathbf{r}^\dagger\Phi_\mathbf{r}=1$, which we shall
only impose on average. From Eq.~(\ref{eq:11}), we require, at zero temperature:
\begin{equation}
\langle\mathsf{s}^+_{\mathbf{r},\mathbf{r}+\mathbf{e}_\mu}\rangle=\frac{1}{2}\frac{\langle\Phi_\mathbf{r}^\dagger\Phi_{\mathbf{r}+\mathbf{e}_\mu}\rangle}{|\langle\Phi_\mathbf{r}^\dagger\Phi_{\mathbf{r}+\mathbf{e}_\mu}\rangle|}=\frac{1}{2}\frac{1}{\langle\Phi_\mathbf{r}^\dagger\Phi_{\mathbf{r}+\mathbf{e}_\mu}\rangle^*}.
\end{equation}
Now, since $H_\Phi$ is quadratic in the $\Phi$'s, we may, at least
formally, write the $\Phi^\dagger\Phi$ Green's functions. First, we
want to impose the constraint
$\langle\Phi_\mathbf{r}^\dagger\Phi_\mathbf{r}\rangle=1$, which we
introduce in the path integral formulation via the term $\lambda(|\Phi_\mathbf{r}|^2-1)$, where $\lambda$ is a Lagrange multiplier.

Let us now assume translational invariance:
\begin{eqnarray}
  H_\Phi&=&-\frac{1}{2}\sum_{\mathbf{r}\in{\rm    A,B}}\sum_\mu h_{\mu}\overline{\mathsf{s}}_{\mu}^{\epsilon_\mathbf{r}}\Phi_\mathbf{r}^\dagger\Phi_{\mathbf{r}+\epsilon_\mathbf{r}\mathbf{e}_\mu}\\
&=&-\frac{1}{2}\sum_{\mathbf{r}\in{\rm A}}\sum_\mu h_{\mu}\overline{\mathsf{s}}_{\mu}^{+}\Phi_\mathbf{r}^\dagger\Phi_{\mathbf{r}+\mathbf{e}_\mu}-\frac{1}{2}
\sum_{\mathbf{r}\in{\rm    B}}\sum_\mu h_{\mu}\overline{\mathsf{s}}_{\mu}^{-}\Phi_\mathbf{r}^\dagger\Phi_{\mathbf{r}-\mathbf{e}_\mu},\nonumber
\end{eqnarray}
where $\overline{\mathsf{s}}_\mu^\pm=\langle
\mathsf{s}_{\mathbf{r},\mathbf{r}+\mathbf{e}_\mu}^\pm\rangle$. Then it is useful to go to
Fourier space and express the Lagrangian:
\begin{eqnarray}
  \mathcal{L}_\Phi&=&
\frac{1}{N_{\rm u.c.}}\sum_\mathbf{k}
\left(\begin{array}{cc}
\Phi_{\mathbf{k},{\rm A}}^\dagger&\Phi_{\mathbf{k},{\rm B}}^\dagger
\end{array}\right)\\
&&\;\cdot\left(\begin{array}{cc}
\lambda+\frac{\omega_n^2}{2J}&-\frac{1}{2}\sum_\mu h_\mu\overline{\mathsf{s}}^+_\mu
e^{i(\mathbf{k}\cdot\mathbf{e}_\mu)}\\
-\frac{1}{2}\sum_\mu h_\mu\overline{\mathsf{s}}^-_\mu
e^{-i(\mathbf{k}\cdot\mathbf{e}_\mu)}&\lambda+\frac{\omega_n^2}{2J}
\end{array}\right)\nonumber\\
&&\quad\cdot\left(\begin{array}{c}
\Phi_{\mathbf{k},{\rm A}}\\\Phi_{\mathbf{k},{\rm B}}
\end{array}\right)
\end{eqnarray}
where the constraint $|\Phi_\mathbf{r}|^2=1$ was relaxed to
$\sum_\mathbf{r}|\Phi_\mathbf{r}|^2=2N_{\rm u.c.}$. So:
\begin{equation}
  G^{-1}=
\left(\begin{array}{cc}
\lambda+\frac{\omega_n^2}{2J}&M_\mathbf{k}\\
M^*_{\mathbf{k}}&\lambda+\frac{\omega_n^2}{2J}
\end{array}\right),\; 
M_{\mathbf{k}}=-\frac{1}{2}\sum_\mu h_\mu\overline{\mathsf{s}}^+_\mu
e^{i(\mathbf{k}\cdot\mathbf{e}_\mu)}.
\end{equation}
Hence,
\begin{equation}
G_{\omega_n,\mathbf{k}}=\frac{1}{\left(\lambda+\frac{\omega_n^2}{2J}\right)^2-|M_\mathbf{k}|^2}
\left(\begin{array}{cc}
\lambda+\frac{\omega_n^2}{2J}&M_\mathbf{k}\\
M_\mathbf{k}^*&\lambda+\frac{\omega_n^2}{2J}
\end{array}\right).
\end{equation}
We are looking for the equal-time correlations, and therefore consider:
\begin{eqnarray}
   &&\frac{1}{2\pi}\int_{-\infty}^{+\infty} d\omega_n\frac{M_\mathbf{k}}{\left(\lambda+\frac{\omega_n^2}{2J}\right)^2-|M_\mathbf{k}|^2}\\
&&=\frac{J M_\mathbf{k}}{2\pi |M_\mathbf{k}|}\\
&&\times\int_{-\infty}^{+\infty}d\omega_n\left(\frac{1}{2J(\lambda-|M_\mathbf{k}|)+\omega_n^2}-\frac{1}{2J(\lambda+|M_\mathbf{k}|)+\omega_n^2}\right)\nonumber\\
&&=\frac{1}{2}\sqrt{\frac{J}{2}}\frac{M_\mathbf{k}}{|M_\mathbf{k}|}\left(\frac{1}{\sqrt{\lambda-|M_\mathbf{k}|}}-\frac{1}{\sqrt{\lambda+|M_\mathbf{k}|}}\right),
\end{eqnarray}
and
\begin{eqnarray}
 &&\frac{1}{2\pi}\int_{-\infty}^{+\infty} d\omega_n\frac{\lambda+\frac{\omega_n^2}{2J}}{\left(\lambda+\frac{\omega_n^2}{2J}\right)^2-|M_\mathbf{k}|^2}\\
&&=\frac{1}{2\pi}\int_{-\infty}^{+\infty} d\omega_n\left[\frac{1}{\lambda+\frac{\omega_n^2}{2J_{zz}}+|M_\mathbf{k}|}+\frac{|M_\mathbf{k}|}{\left(\lambda+\frac{\omega_n^2}{2J}\right)^2-|M_\mathbf{k}|^2}\right]\nonumber\\
&&=\frac{1}{2}\sqrt{\frac{J}{2}}\left(\frac{1}{\sqrt{\lambda-|M_\mathbf{k}|}}+\frac{1}{\sqrt{\lambda+|M_\mathbf{k}|}}\right).
\end{eqnarray}
Clearly, $\lambda\geq\lambda_{\rm min}=\max_\mathbf{k}|M_\mathbf{k}|$. For $\overline{\mathsf{s}}_\mu^+=1/2$ and $h_\mu=h$, $\lambda_{\rm min}=h$.

The condition 
$\sum_{\mathbf{r}\in{\rm A}}\langle\Phi_\mathbf{r}^\dagger\Phi_\mathbf{r}\rangle=N_{\rm
u.c.}$ yields:
\begin{equation}
\label{eq:14}
1=\frac{1}{2}\sqrt{\frac{J}{2}}\int_\mathbf{k}\left(\frac{1}{\sqrt{\lambda-|M_\mathbf{k}|}}+\frac{1}{\sqrt{\lambda+|M_\mathbf{k}|}}\right).
\end{equation}
The second order transition to the state connected to the high-``field'' state is given by the solutions to Eq.~(\ref{eq:14}) taken at
$\lambda=\lambda_{\rm min}$. At uniform $h_\mu=h$, with $\overline{\mathsf{s}}^+_\mu=1/2$, we obtain $(h/J)_c\approx0.35$, as given in the main text.

\end{document}